%% file: main.tex
\documentclass[sigconf]{acmart}
\usepackage{booktabs} 
\usepackage[]{todonotes}
\usepackage{color}
\usepackage{amsmath}
\usepackage{amssymb}
\usepackage[T1]{fontenc}
\usepackage[utf8]{inputenc}
\usepackage{hyperref}
\usepackage{tabularx}
\usepackage{multirow}

\usepackage{listings} 
\usepackage{newfloat}
\usepackage[inline]{enumitem}

\DeclareFloatingEnvironment[
fileext=ltt,
listname={List of Listings},
name=Listing
]{listing}

\newcommand\setrow[1]{\gdef\rowmac{#1}#1\ignorespaces}
\newcommand\clearrow{\global\let\rowmac\relax}
\clearrow

\newenvironment{myquote}[1]%
{\list{}{\leftmargin=#1\rightmargin=#1}\item[]}%
{\endlist}

\begin{document}

\title{Inspection Guidelines to Identify Security Design Flaws}

\author{Katja Tuma, Danial Hosseini, Kyriakos Malamas, Riccardo Scandariato}
\affiliation{%
	\institution{Chalmers | University of Gothenburg}
	\city{Gothenburg}
	\country{Sweden}}
\email{katja.tuma@cse.gu.se, daniel.hosseini1@gmail.com, kmalamas@outlook.com, riccardo.scandariato@cse.gu.se}

%
%

\begin{abstract}
Recent trends in the software development practices (Agile, DevOps, CI) have shortened the development life-cycle causing the need for efficient security-by-design approaches. 
In this context, software architectures are analyzed for potential vulnerabilities and design flaws.
Yet, design flaws are often documented with natural language and require a manual analysis, which is inefficient.
Besides low-level vulnerability databases (e.g., CWE, CAPEC) there is little systematized knowledge on security design flaws.
The purpose of this work is to provide a catalog of security design flaws and to empirically evaluate the inspection guidelines for detecting security design flaws.
To this aim, we present a catalog of 19 security design flaws and conduct empirical studies with master and doctoral students.
This paper contributes with: (i) a catalog of security design flaws, (ii) an empirical evaluation of the inspection guidelines with master students, and (iii) a replicated evaluation with doctoral students.
We also account for the shortcomings of the inspection guidelines and make suggestions for their improvement with respect to the generalization of guidelines, catalog re-organization, and format of documentation.
We record similar precision, recall, and productivity in both empirical studies and discuss the potential for automating the security design flaw detection.
\end{abstract}


\maketitle

\input{sec/introduction}
\input{sec/catalog}
\input{sec/experiment}
\input{sec/results}
\input{sec/improvements}
\input{sec/relatedwork}
\input{sec/threats}
\input{sec/conclusions}


\bibliographystyle{ACM-Reference-Format}
\bibliography{lit}


\end{document}

%% file: sec/introduction.tex
\section{Introduction}
\label{sec:intro}
Recent trends in software development, such as, Agile, DevOps, and Continuous Integration (CI), have shortened the software development life-cycle, impacting software security \cite{mcgraw2017six,ebert2017industry}.
For instance, CI tightened release cycles to days, or sometimes hours.
This limits the activities that can take place for security analysis, causing the need for \textit{efficient} security-by-design approaches.
In the design phase of the development life-cycle, software architectures are often analyzed for potential design flaws and vulnerabilities.
Knowledge reuse is an important factor that can help raise the efficiency.
For instance, previous work (\cite{berger2016automatically,wang2010ranking,almorsy2013automated}, to cite a few) has made use of publicly available records of low-level security vulnerabilities, such as CAPEC \footnote{\url{https://capec.mitre.org}}, CVE \footnote{\url{https://www.cvedetails.com}}, CWE \footnote{\url{https://cwe.mitre.org}} to semi-automate the security analysis of systems.
On the level of software architecture, Garcia et al. \cite{garcia2009toward} introduce a catalog of architectural bad smells specified with UML diagrams.
Similarly, Bouhours et al. \cite{bouhours2009bad} contribute with a catalog of 23 so called ``spoiled patterns'' or, architectural design antipatterns.
Yet, the existing literature about architectural design flaws \cite{garcia2009toward,bouhours2009bad,taibi2018definition,mo2019architecture,nafees2017vulnerability} lacks a systematized knowledge about security-relevant architectural design flaws.
In addition, there is a lack of practical inspection guidelines for identifying security design flaws in software architectures.

This paper contributes with: 
(i) inspection guidelines accompanied by a series of closed questions for detecting 19 security design flaws,
(ii) an experiment with master students investigating precision, recall, and productivity of the guidelines, and
(iii) a replicated experiment with doctoral students.
In the replicated experiment we track the problematic guidelines and suggest improvements to the catalog with respect to: (a) generalizing the guidelines, b) re-organizing the catalog, and c) format of documentation.

We record a relatively high precision ($92.6 \%$) and productivity ($11.5 \ TP/h$).
On the other hand, our results show that about half of the security design flaws go unnoticed (average recall is $50.4\%$).
Similar measurements of precision and recall have been reported in related empirical studies investigating knowledge-based manual threat analysis techniques, i.e. STRIDE \cite{tuma2018two,scandariato2015descriptive}.
We have found that many participants have expressed doubts when detecting certain security design flaws.
We systematically track which guidelines were problematic and provide an account for the re-occurring issues.
Accordingly, we suggest simple improvements to overcome these issues.
The detection guidelines can be currently used to manually analyze software design.
However, we see potential in using these guidelines to automate the detection of security design flaws.
Specifically, the closed questions are already operationalized to some extent, and can be easily transformed into graph queries. 

The rest of the paper is organized as follows.
Section \ref{sec:catalog} describes the guidelines for detecting security design flaws and shows how they are used. 
Section \ref{sec:experiment} describes the experimental design and execution.
Section \ref{sec:results} presents the results and 
Section \ref{sec:improve} suggests improvements for the guidelines.
Section \ref{sec:threats} lists the threats to validity, while 
Section \ref{sec:relatedwork} discusses the related work. We conclude the paper in
Section \ref{sec:conclusion}.


%% file: sec/catalog.tex
\section{Security Design Flaws Catalog}
\label{sec:catalog}
This section gives a short description of the developed security design flaws catalog.
We refer the interested reader for a complete catalog to \cite{hosmal2017secdfcatalog}.

\begin{table}
	\center
	\footnotesize{}
	\caption{A list of the proposed security design flaws.}
	\label{tab:catalogsummary}
	\setlength\tabcolsep{2.0pt}
	\begin{tabular}{ p{3cm} p{5cm}}
		\toprule
		 \multicolumn{1}{c}{Name} & \multicolumn{1}{c}{Description} \\
		\midrule
		Missing authentication & An absence of an auth. mechanism in the system. \\
		Authentication bypass & The auth. mechanism does not cover all possible entry points to the system. \\
		Relying on single factor auth.  & The auth. mechanisms rely on the use of passwords. \\
		Insuff. session management & Sessions are not managed securely throughout their life-cycle. \\
		Downgrade authentication  & Possibility to authenticate with a weaker (or obsolete) auth mechanism. \\
		Insuff. crypto key management  & Keys are not managed securely throughout their life-cycle. \\
		Missing authorization  & An absence of an authorization mechanism in the system. \\
		Missing access control  & An absence of access control in the system. \\
		No Re-authentication  &  An absence of re-authentication during critical operations. \\
		Unmonitored execution & Uncontrolled resource consumption due to interactions with external entities. \\
		No context when authorizing & An absence of conditional checks for access control. \\
		Not revoking authorization & An absence of a process for revoking user access. \\
		Insecure data storage & Storage of sensitive data is in clear or weak access control mechanisms are in place. \\
		Insuff. credentials management  & Credentials are not managed securely throughout their life-cycle. \\
		Insecure data exposure & Sensitive data is transported in clear text. \\
		Use of custom/weak encryption & Generating small keys, using obsolete encryption schemes.   \\	
		Not validating input/data & Absence of validation checks when receiving data from external entities. \\
		Insuff. auditing & Access to critical resources or operations is not logged. \\
		Uncontrolled resource consumption & Uncontrolled resource consumption of internal components. \\
		\bottomrule
	\end{tabular}
\end{table}
Table \ref{tab:catalogsummary} shows the security design flaws and their descriptions.
The catalog is a list of 19 design flaws related to issues with authentication, access control, authorization, availability of resources, integrity and confidentiality of data.
The catalog entries consist of (a) the name of the design flaw, (b) a description (using natural language), and (c) a series of closed questions that serve as detection guidelines.
\begin{listing}\centering{}
	\fbox{\parbox{0.95\columnwidth}{
			\begin{center}
				Design Flaw 1: Missing authentication
			\end{center}
			\footnotesize{}
			\begin{description}
				\item[Description] This refers to the absence of an authentication mechanism in the system. Apart from external entities, like users or other systems the system may interact with, authentication may be necessary within the system between processes/components/datastores that are located in different trust boundaries.
				\item[Detection] ~
				\begin{itemize}[leftmargin=0cm]
					\item Consider the external entities (users/subsystems) that interact with the system and which assets of the system they can access.
					\item Determine the processes that interact with high-value assets in the system.
					\item For each interaction examine:
					\begin{itemize}
						\item If it is an entity: Does the entity go through an authentication point in order to access the asset?
						\item If it is a process: Is the identity of a process accessing datastores or processes in a different part of the system (trust boundaries – requires different privilege levels) verified?
					\end{itemize}
				\end{itemize}
			\end{description}
	}}
	\caption[Design Flaw 1: Missing authentication]{\label{lst:missing-auth} Textual description of the missing authentication design flaw.}
\end{listing}
Listing \ref{lst:missing-auth} shows the first catalog entry.
The catalog was compiled by systematically filtering vulnerability database entries (CVE, CWE, OWASP\footnote{\url{https://www.owasp.org/index.php/Category:OWASP_Top_Ten_Project}}, and SANS\footnote{\url{https://www.sans.org/top25-software-errors/}}) and existing threat and vulnerability taxonomies.
The final catalog entries were obtained by grouping a filtered subset of database entries and taxonomies.
The authors grouped the entries whenever the vulnerabilities could be caused by the same design-level issues.
For instance, they relate 3 CWE entries ($287, 306, $ and $862$) to the security design flaw ``Missing authentication'' (Listing \ref{lst:missing-auth}).
The CWE entry 287 is a description of an improper authentication, where the software is not able to prove the identify of the actors in the system.
Entry 306 is a description of commonly missing (re)authentication mechanisms when critical functions are executed.
Finally the entry 862 is describing the weakness of missing authorization mechanisms system actors communicating with the system.
All the above entries are describing weaknesses related to external entities or critical processes communicating with the system without an authentication mechanism in place.
Therefore, the design issue causing all three entries is missing authentication.
A detailed procedure of the catalog compilation can be found in \cite{hosmal2017secdfcatalog}.

%% file: sec/experiment.tex
\section{Empirical Experiments}
\label{sec:experiment}

\emph{Research questions.}
We conducted two controlled experiments with participants to investigate the efficiency and effectiveness of using existing guidelines for detecting security design flaws in a software architecture.
The first experiment was conducted by the second and third author with master students.
The second experiment was conducted by the first author with doctoral students.
We were interested to measure the efficiency and effectiveness of the guidelines in both experiments.
We observe effectiveness of the guidelines with a measure of \textit{precision} and \textit{recall}, and efficiency with a measure of \textit{productivity}.
This study focuses on the following research questions.

\noindent
\textbf{RQ1.}
What is the precision, recall, and productivity of the proposed guidelines for security design flaw detection?

\noindent
\textbf{RQ2.}
What are the shortcomings and benefits of the proposed guidelines for security design flaw detection?

\smallbreak
\emph{Experimental object.}
\begin{figure}
	\centering
	\includegraphics[width=1.0\columnwidth]{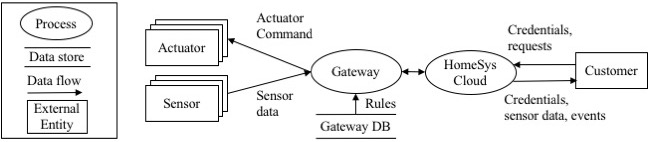}
	\caption{A high-level DFD of the experimental object.}
	\label{fig:contextDFD}
\end{figure}
Figure \ref{fig:contextDFD} shows the context DFD of the experimental object.
The Home Monitoring System (HomeSys) is a system for remotely monitoring private residents. 
Its purpose is to provide the infrastructure and functionalities for customers to automatically receive and manage notifications about critical events in their homes.
In principle, the system consists of a gateway communicating with sensors, and a cloud system collecting data from gateways and displaying it on customer dashboards.
Sensors are analog or digital hardware devices that produce measurements and send them to the gateway.
Actuators are hardware devices that can receive commands from the gateway, like for instance, taking a picture, activating a buzzer or flicking a switch.
The gateway is a hardware device which relays measurements to the cloud (via a 3G or WiFi network) and manages the actuators in the residency.
The cloud is a software system which communicates with the gateways and provides services for the customers.

The system documentation (about 30 pages) includes (1) the description of the problem domain with scenarios, (2) the requirements of the system including non-functional requirements and (3) a hierarchically decomposed architecture specified with UML (e.g, deployment diagram).
The complete description of the system has been used in previous studies \cite{tuma2018two} and is open to the public.
The participants were tasked to read the preparatory material (including the system documentation) before the experiment took place.
Email reminders were sent out before-hand.
The documentation of the system does not include security requirements, as the purpose of this exercise is to derive those from the detected design flaws.
\smallbreak
\emph{Participants.}
The participants of this study are three master-level students and three doctoral candidates in computer science and engineering disciplines.
We excluded one report handed in by a master student as the task was not taken seriously.
All the participants had experience with modeling, basic security concepts, and UML notations used in the system documentation.
The participants in the second experiment were doctoral candidates in software engineering with a master degree in computer science and engineering.
They have worked with modeling before, and were familiar with basic security concepts and the notations used for describing the experimental object.
We have made sure that the participants were able to complete the task before starting the experiment.
The participants performed the task individually. 
\smallbreak
\emph{Task.}
In essence, the participants were tasked to go through each entry in the catalog, use the guidelines to detect flaws in the HomeSys design, and document the identified flaws.
The guidelines prescribe detecting a subset of elements in the architecture for inspection, therefore the participants had to also indicate the location of the identified flaw.
For each architectural element under inspection, the guidelines provide a list of closed questions.
A negative answer indicates the existence of a design flaw for that element.  
In case of insufficient information in the documentation, the participants were instructed to report a design flaw, anyway.
If they thought a particular entry was not applicable in the system under examination due to some restrictions or regulations in the domain, they were instructed to mark that entry as not applicable. 
Finally, the participants handed in all the printed documents, including the filled-in form documenting the detected flaws.
\smallbreak
\emph{Execution.}
Before the execution of the first experiment, four participants were selected due to having successfully completed a master level course on advanced software architecture.
Therefore, they were familiar with basic concepts of software architecture design, and have studied analyzing architectures for security in the context of this course.
On the day of the first experiment, the participants were gathered in a classroom.
They were given printed copies of (i) a one-page task description (ii) the HomeSys documentation, (iii) the catalog of security design flaws as described in Section \ref{sec:catalog}, and (iv) a form for documenting the identified flaws.
Regular breaks were allowed during which the experimenters made sure the participants did not compare solutions.

Before the execution of the second experiment, preparatory reading was handed out via electronic mail (2 weeks in advance).
The preparatory reading included the same documents used in the first experiment.
Regular email reminders were sent until the last day before the execution.
On the day of the experiment, the participants met the experimenter for an individual session on faculty premises.
They was given printed copies of all the documents (i-iv).
The first author explained their task again and briefly described the printed documents.
Only procedural questions were answered during the experiment.
Due to the complexity of the task, no strict time limit was enforced, short breaks were permitted, and accounted for.

\smallbreak
\emph{Measures.}
We adopt the same ground truth and measures of precision, recall, and productivity in both experiments.
The ground truth was re-assessed after the first experiment to ensure it's correctness.
It consists of 48 security design flaws.
Conventionally, precision $(TP/(TP+FP))$ is measured as a ratio between the true positives (i.e., correctly identified flaws) and all identified flaws (including the false flaws).
A true positive $(TP)$ is a correctly identified security design flaw.
This entails that the documented flaw exists also in the ground truth and is identified at the same location of the architecture.
A false positive $(FP)$ is an incorrectly identified and documented flaw that does not exist in the ground truth.
Recall $(TP/(TP+FN))$ is measured as a ratio between the true positives and all correctly identified flaws (including the overlooked flaws).
A false negative $(FN)$ is a design flaw that exists in the ground truth but has not been documented.
Productivity $(TP/h)$ is measured as the amount of correctly identified flaws per hour.
We measured the time it took for the participants to complete the task and subtract the time that was lost during the breaks.
We include an additional measure for keeping track of the guidelines the participants struggled with.
We flag the $TP$, $FP$, $FN$ whenever the participant expressed that there was insufficient information ($II$) to determine the existence of a flaw.
If a correctly identified flaw $(TP)$ is flagged with $II$ this means that the participant has identified the correct flaw in the correct location in the architecture, has documented this in the hand-in, but has also expressed doubt due to missing information in the documentation.
If an overlooked flaw $(FN)$ is flagged with $II$ this means that the participant failed to identify the correct flaw but also expressed doubt due to missing information.

%% file: sec/results.tex
\section{Results}
\label{sec:results}

\begin{table}
	\center
	\footnotesize{}
	\caption{Results from both experiments (accumulated result are in bold).}
	\label{tab:results}
	\begin{tabular}{>{\rowmac}l >{\rowmac}l >{\rowmac}r >{\rowmac}r >{\rowmac}r >{\rowmac}r >{\rowmac}r >{\rowmac}r<{\clearrow}}
		\toprule
		 & & \multicolumn{3}{c}{Design Flaws} & \multicolumn{3}{c}{Measure}\\
		\cmidrule(lr){3-5}\cmidrule(lr){6-8}
		& Participant ID & \multicolumn{1}{c}{TP} & \multicolumn{1}{c}{FP} & \multicolumn{1}{c}{FN} & \multicolumn{1}{c}{P $[\%]$} & \multicolumn{1}{c}{R $[\%]$} & \multicolumn{1}{c}{Prod $[TP/h]$}  \\
		\midrule
		\parbox[t]{2mm}{\multirow{4}{*}{\rotatebox[origin=c]{90}{MSc}}} & 1 & 23 & 2 & 24 & 92 & 48.9 & 9.5 \\
		& 2 & 26 & 3 & 21 & 89.7 & 55.3 & 9.4 \\
		& 3 & 19 & 4 & 28 & 82.6 & 40.4 & 6 \\
		\cmidrule(lr){2-8}
		& Avg & 22.7& 3 & 24.3 & 88.1 & 48.2 &  8.3 \\
		\midrule
		\parbox[t]{2mm}{\multirow{4}{*}{\rotatebox[origin=c]{90}{PhD}}} & 4 & 24 & 1 & 23 & 96 & 51.1 & 8.2 \\
		& 5 & 20 & 1 & 27 & 95.2 & 42.6 & 8.5 \\
		& 6 & 30 & 0 & 17 & 100 & 63.8 & 17.7 \\
		\cmidrule(lr){2-8}
		& Avg & 24.7 & 0.7 & 22.3 & 97.1 & 52.5 & 14.8 \\
		\midrule
			\setrow{\bfseries} & Total Avg & 23.7 & 1.8 & 23.3 & 92.6 & 50.4 & 11.5  \\
		\bottomrule
	\end{tabular}
\end{table}
Table \ref{tab:results} summarizes the results of both experiments.
First, we have calculated the average precision, recall, and productivity for both experiments separately.
On average, participants 4-6 from Table \ref{tab:results} performed slightly better compared to participants 1-3 (avg $97.1\%$ precision vs. $88.1\%$, avg $52.5\%$ recall vs. $48.2\%$, $14.8 \ TP/h$ productivity vs. $8.3 \ TP/h$).
Yet, these differences are small and not significant.
The performance differences between experiments can be explained by the different level of education.
Henceforth we refer to the accumulated $TP$, $FP$, $FN$, measures of precision, recall, and productivity of the inspection guidelines.

\smallbreak
\emph{Precision, recall, and productivity.}
On average, the participants identified about half $(23.7/48)$ of the security design flaws correctly.
Yet, on average only about 2 reported security design flaws were incorrect $(FP)$.
Therefore, the average \textit{precision} is quite high (\textbf{92.6 \%}).
The low number of $FPs$ may indicate that the inspection guidelines were not misleading the participants towards a false design flaw discovery.
On the other hand, about half of the flaws were overlooked.
On average, the \textit{recall} is measured at \textbf{50.4 \%}.
This result is not surprising.
Similar measurement of precision and recall have been reported in related empirical studies investigating manual knowledge-based threat analysis techniques, i.e., STRIDE \cite{tuma2018two,scandariato2015descriptive}.
In general, high precision and low recall may be a common trait for techniques that manually analyze software architectures.
Finally, on average the participants spent about 2.5 hours to complete the task.
The average \textit{productivity} of the approach is \textbf{11.5} correct threats per hour.
This result is more surprising, as the related literature reports a much lower number of correctly identified threats per hour ($1.8 \ TP/h$ in \cite{scandariato2015descriptive} and about $4 \ TP/h$ in \cite{tuma2018two}).
This can be explained by the different goals of threat analysis techniques vs detection of design flaws.
Threat analysis techniques help to systematically identify security threats on the level of software architecture.
The security threats are considered correct only when a realistic attack scenario is found.
Finding a realistic scenario requires thinking about possible attack paths and how to break into the system.
This may be cognitively more demanding than answering a set of closed questions about the architectural design.



%

\smallbreak
\emph{Insufficient Information.}
In the second experiment, we measure which guidelines posed problems to our participants.
Particularly, we flag correctly identified and overlooked flaws with \emph{Insufficient Information (II)}.
To this aim, we have re-assessed the data collected in both experiments.
A $FN$ was flagged when the participant reported the flaw, but never specified the location due to missing information.
A $FN$ was also flagged when the participant did not report the flaw, but made notes about missing information next to the check-list in the catalog.
Incorrectly identified flaws $(FP)$ were never flagged.
The participants never expressed doubt about missing information when making a mistake.

\begin{table}
	\center
	\footnotesize{}
	\caption{Flaws flagged with insufficient information $(II)$ (problematic flaws are in bold).}
	\label{tab:resultsII}
	\begin{tabular}{>{\rowmac}l  >{\rowmac}r  >{\rowmac}r >{\rowmac}r <{\clearrow}}
		\toprule
		Flaw ID & \multicolumn{1}{c}{TP (Participant ID)} & \multicolumn{1}{c}{FN (Participant ID)} & \multicolumn{1}{c}{$\Sigma$} \\
		\midrule
		2 & 4 (6) & 0 & 4 \\
		3 & 1 (6) & 0 & 1 \\
		\setrow{\bfseries}4 & 7 (1,2,3,6) & 3 (3,5) & 10 \\
		7 & 0 & 4 (5) & 4 \\
		9 & 8 (1,2) & 0 & 8 \\
		\setrow{\bfseries}12 & 3 (1,2) & 11 (1,2,5,6) & 14 \\
		13 & 2 (6) & 2 (6) & 4 \\
		14 & 2 (5,6) & 0 & 2 \\
		\setrow{\bfseries}15 & 1 (6) & 15 (1,2,5,6)  & 16 \\
		17 & 2 (5) & 2 (6) & 4 \\
		\setrow{\bfseries}18 & 0 & 2 (3,4) & 2 \\
		19 & 2 (6) & 0 & 2 \\
		\midrule
		Total & 32 & 39 & 71\\
		\bottomrule
	\end{tabular}
\end{table}
Table \ref{tab:resultsII} shows the flagged $TP$, $FP$, which flaw they relate to, and how many participants expressed the same doubts.
We have recorded that $\textbf{25.2\%} \ (71/282)$ of all reported TP and FN from Table \ref{tab:results} $(\Sigma TP + \Sigma FN = 282)$ were flagged with $II$.
Visibly, security design flaws 4, 12, 15, and 18 seem to be problematic.
These design flaws were often subject to missing information, for more than one participant.
This may indicate that some guidelines assume the availability of detailed information about the system (information that is rarely available in the design phase).
In addition, we have gathered participants feedback in the exit questionnaire.
We asked the participants about what they did not like about the approach.
Some participants referred to missing information. For instance:
\begin{myquote}{0.1in}
\textit{``I felt that the approach and the architecture were too detached, and that I \textbf{needed much more information} that what was provided in the architecture to complete the analysis properly.''}
\end{myquote}
This confirms that the participants indeed had problems with some of the guidelines.
In particular we have identified three kind of problems, and suggest improvements in what follows.


%% file: sec/improvements.tex
\section{Suggestions for improving the inspection guidelines}
\label{sec:improve}
In what follows we describe three type of problems we encountered with the proposed guidelines, and provide suggestions for improvement.
\smallbreak
\emph{Guidelines generalization.}
We analyze the problematic guidelines related to design flaws 4 and 15.
Both guidelines are \textbf{not general enough} to be useful for a design-level analysis.
We have observed this trend in other guidelines as well (namely, 12, 18).

\begin{listing}\centering{}
	\fbox{\parbox{0.95\columnwidth}{
			\begin{center}
				Design Flaw 4: Insufficient Session Management
			\end{center}
			\footnotesize{}
			\begin{description}
				\item[Description] Not managing a session properly throughout its life-cycle can leave the system vulnerable to session hijacking attacks. Session management involves creation (the session should be established through a secure channel and session identifier should be encrypted), the time frame the session is active (an attacker might attempt to reuse the session ID to gain access) and its destruction/invalidation (proper session invalidation should take place when the user logs out or session timeout. Not terminating sessions can also lead to resource depletion).
				\item[Detection] ~
				\begin{itemize}[leftmargin=0cm]
					\item Determine which sessions are established in the system and between which endpoints.
					\item For each session examine:
					\begin{itemize}
						\item Is the session established through a secure channel?
						\item Is the session ID encrypted when in transit?
						\item Is the session ID hard to guess?
						\item Is the use of session ID as a parameter in URLs prevented?
						\item Is the session ID validated on server side?
						\item Are secure cookies used?
						\item Is the session ID tied to other user properties like IP, SSL session ID?
						\item Can the same session be accessed simultaneously from two endpoints? Should it?
						\item Is session timeout set? Is it the minimum possible value?
						\item Is the session invalidated on logout?
						\item Is the session ID renewed in the event of privilege change?
						\item Is there a mechanism to monitor the creation/destruction and attempts to connect to a session?
						\item Is the user required to re-authenticate after a period of inactivity?
					\end{itemize}
				\end{itemize}
			\end{description}
	}}
	\caption[Design Flaw 4: Insecure Data Exposure]{\label{lst:txt-session-management}Textual description of the insufficient session management design flaw.}
\end{listing}

Listing \ref{lst:txt-session-management} shows the guidelines related to security design flaw 4.
Overall, participants still managed to correctly identify this flaw despite having doubts due to missing information (c.f., Table \ref{tab:resultsII}).
They were able to do so as some question posed for detection are in fact very useful (e.g., \textit{Is the user required to re-authenticate after a period of time?}) and can be answered at design time.
One possible explanation for our participants expressing doubts is that there were too many technical questions posed for detection.
For instance, \textit{Is the session ID tied to other properties like IP, SSL,?}
The properties of the session IDs are technology-dependent.
The choice of technology is decided in later stages of the development life-cycle \cite{medvidovic2010software}.

\begin{listing}\centering{}
	\fbox{\parbox{0.95\columnwidth}{
			\begin{center}
				Design Flaw 15: Insecure Data Exposure
			\end{center}
			\footnotesize{}
			\begin{description}
				\item[Description] Data is not transferred in a secure way. For example a web application uses the HTTP instead of HTTPS\@.
				This leaves the channel vulnerable to eavesdropping, Man In The Middle (MITM) attacks etc.
				\item[Detection] ~
				\begin{itemize}[leftmargin=0cm]
					\item Locate the valuable information in the model. 
					\item Track them through the architecture to determine where and how they are transferred. 
					\item At each step examine the following:
					\begin{itemize}
						\item Is the reuse of packets prevented (Replay attacks)?
						\item Is there any form of timestamping, message sequencing or checksum in the exchanged packages?
						\item Is the traffic over an encrypted channel (SSL/TLS)?
					\end{itemize}
				\end{itemize}
			\end{description}
	}}
	\caption[Design Flaw 15: Insecure Data Exposure]{\label{lst:txt-insecure-data-exposure}Textual description of the insecure data exposure design flaw.}
\end{listing}
Listing \ref{lst:txt-insecure-data-exposure} shows the guidelines related to security design flaw 15.
Compared to the previous flaw, these guidelines are much shorter.
All three questions posed for flaw detection are very technical.
For instance, \textit{Is the reuse of packets prevented (Replay attacks)?}
The participants did not know about replay attacks are, or how to counter them.
This may have caused the participants to simply ignore this flaw without even finding the possible locations, and making a note about missing information (resulting in a flagged $FN$).

\begin{listing}\centering{}
	\fbox{\parbox{0.95\columnwidth}{
			\begin{center}
				Design Flaw 4.1: Insufficient Session Management
			\end{center}
			\footnotesize{}
			\begin{description}
				\item[Description] Not managing a session properly throughout its life-cycle can leave the system vulnerable to session hijacking attacks. Session management involves the creation, the time frame the session is active, and its destruction. The attacker may attempt to disrupt or manipulate these processes for her gain.
				\item[Detection] ~
				\begin{itemize}[leftmargin=0cm]
					\item Determine which sessions are established in the system and between which endpoints.
					\item For each session examine:
					\begin{itemize}
						\item Is the session established through a secure channel?
						\item Is the session ID hard to guess?
						\item Is the session ID protected when in transit (e.g., encrypted)?
						\item Is there a process for validating the session ID on server side?
						\item Is the session destructed (or invalidated) on logout?
						\item Is the session ID renewed in the event of privilege change?
						\item Is the user required to re-authenticate after a period of inactivity?
					\end{itemize}
				\end{itemize}
			\end{description}
	}}
	\caption[Design Flaw 4.1: Insecure Data Exposure]{\label{lst:txt-session-management1}Textual description of the improved insufficient session management design flaw.}
\end{listing}
Listing \ref{lst:txt-session-management1} introduces a few simple improvements.
First, the description of the design flaw is shortened and redundant questions removed.
For example, session timeout directly relates to the last posed question \textit{Is the user required to re-authenticate after a period of inactivity?}
Second, the questions posed for detection that were too specific were removed.
For instance, the usage of secure cooking is technology-dependent.

This kind of improvement could help to relate the guidelines to a design-level description, making them easier to apply.
In the exit questionnaire we also ask the participants for their suggestions for improvement.
Generally, the participants suggested similar improvements.
One participant even suggested to have several versions of the same guideline but on different level of abstraction:
\begin{myquote}{0.1in}
\textit{``Maybe divide the 19 points [Security Design Flaws] into classes based on what \textbf{level of design} is provided.''}
\end{myquote}

\smallbreak
\emph{Overlapping guidelines.}
Most participants agreed that the approach takes too much time.
We found that many guidelines are \textbf{overlapping} and could be compressed and re-organized.
In addition, as per comparing guidelines in Listing \ref{lst:txt-session-management} and Listing \ref{lst:txt-insecure-data-exposure}, the guidelines are not well balanced in length and complexity. 
For instance, the guidelines in Listing \ref{lst:txt-insecure-data-exposure} try to help detecting where data can be leaked in transit.
Yet, the proposed catalog contains another set of guidelines to detect the design flaw for insecure data storage (Flaw 13).
Detecting both flaws requires to first identify valuable information and track it through the architecture.
Therefore, merging them and removing redundant questions would help speed up the manual detection.
The exit questionnaire revealed that the participants generally agree with this notion.
For instance, two participants criticized the categorization of guidelines:
\begin{myquote}{0.1in}
	\textit{``Maybe I would not categorize them [Security Design Flaws 1-19] so much. Having 5-10 [questions] under the same category \textbf{does not always represents the category}.''}
\end{myquote}
\begin{myquote}{0.1in}
	\textit{``Try to group the items of the list (e.g., there are many references to \textbf{encryption scattered through all the guidelines}).''}
\end{myquote}

\smallbreak
\begin{table}
	\center
	\footnotesize{}
	\caption{Suggestion for re-organization of the proposed guidelines.}
	\label{tab:reorg}
	\setlength\tabcolsep{2.0pt}
	\begin{tabular}{l p{4cm} l p{3cm}}
		\toprule
		\multicolumn{2}{c}{Old} & \multicolumn{2}{c}{New}\\
		\cmidrule(lr){1-2}\cmidrule(lr){3-4}
		ID & \multicolumn{1}{c}{Name} & \multicolumn{1}{c}{ID} & \multicolumn{1}{c}{Name} \\
		\midrule
		1 & Missing authentication & 1 & Missing authentication \\
		2 & Authentication bypass &  &  \\
		\midrule
		3 & Relying on single factor auth & 2 & Weak authentication \\
		5 & Downgrade authentication &  &  \\
		9 & No Re-authentication &  &  \\
		14 & Insuff. credentials management &  &  \\
		\midrule
		6 & Insuff. crypto key management & 3 & Missing/weak encryption \\
		16 & Use of custom/weak encryption &  & \\	
		\midrule
		7 & Missing authorization & 4 & Missing authorization \\
		8 & Missing access control &  &  \\
		\midrule
		11 & No context when authorizing & 5 & Weak authorization\\
		12 & Not revoking authorization & & \\
		\midrule
		13 & Insecure data storage & 6 & Leaking important data \\
		15 & Insecure data exposure & & \\
		\midrule
		10 & Unmonitored execution & 7 & Uncontrolled resources \\
		19 & Uncontrolled resource consumption & & \\
		\midrule
		17  & Not validating input/data & - & \\
		18  & Insuff. auditing & 8 & Insuff. auditing\\
		4 & Insuff. session management & 9 & Insuff. session management \\
		\bottomrule
	\end{tabular}
\end{table}
\noindent
Table \ref{tab:reorg} suggests a reorganization of the proposed guidelines.
Implementing such a restructuring (incl. the removal of redundant guidelines) could result in half the original size of the catalog (9 vs 19).
A closer look into the guidelines showed that many security design flaws were very similar.
For instance, the guidelines for detecting missing authorization and missing access control were asking the user for the same kind of information twice.
We also suggest to gather all encryption-related guidelines and group them into one comprehensive guideline (\textit{Missing/weak encryption}).
We suggest the removal of one security design flaw altogether, namely \textit{Not validating input/data} as missing field validation is a bad programming practice, rather than a design flaw. 

\smallbreak
\emph{Tedious documentation.}
Finally, in the exit-questionnaire, participants have expressed a dislike for navigating through documents and documenting the flaws.
For instance, when asked if the approach takes too much time, one participant responded:
\begin{myquote}{0.1in}
\textit{``Yes, especially since you have to go through all the different documents.''}
\end{myquote}
Another participant responded:
\begin{myquote}{0.1in}
\textit{``Yes, some of the questions covered many parts of the system which lead to a journey on finding information.''}
\end{myquote}
The participants had to answer not only the closed questions next to the guidelines, but also fill-in a form with a list of identified flaws and their description.
This resulted in a waste of time and could have only decreased the level of concentration.

\smallbreak
First, for future studies, we suggest to minimize the amount of different documents handed out to participants.
Second, we suggest minor changes in the format of the guidelines.
Along-side each closed question, there should be (i) an obligatory field to specify the location in the architecture, (ii) optional filed for marking a flaw as not applicable, (iii) optional field for notes.
Finally, we suggest to remove the additional form with hand-written descriptions of identified design flaws.
Having these changes in place would mean the users only work with two documents: one describing the architecture, and one with the guidelines and identified flaws.
Automation and tool support would also help reduce this problem.


%% file: sec/relatedwork.tex
\section{Related Work}
\label{sec:relatedwork}
This section positions the paper in the context of existing literature on catalogs of design flaws, vulnerabilities, architectural bad smells, anti-patterns, and knowledge-based threat analysis techniques.

\emph{Catalogs of security design flaws.}
Da Silva and Cecilia \cite{da2016toward} have compiled a catalog of common architectural weaknesses (Common Architectural Weakness Enumeration, CAWE). 
The authors identify and categorize common types of vulnerabilities rooted in software architecture design and provide mitigations to address such vulnerabilities.
Da Silva and Cecilia \cite{da2016toward} also analyze the vulnerabilities of four real systems to discover their cause and find that up to $35\%$ vulnerabilities were rooted in the architectural design.
Similar to our work, the authors investigate which security patterns are likely to have associated vulnerabilities.
However, the proposed catalog is not evaluated with an empirical study.
As a result of initiatives launched by  IEEE Computer Society, Arce et al. \cite{arce2014avoiding} compiled a list of top 10 security design flaws and discuss how to avoid them.
The practical examples that showcase the flaws are very useful for understanding the impacts of each flaw.
Some catalog entires by Hosseini and Malamas \cite{hosmal2017secdfcatalog} are aligned to top 10 security design flaws.
For instance, Flaw 6 (insufficient cryptographic keys management) in \cite{hosmal2017secdfcatalog} relates to the ``use cryptography correctly'' flaw.
Yet, the purpose of the top 10 security design flaws was to raise awareness among software architects about the most common issues that have been the leading cause for security breaches in practice.
The purpose of this paper was to re-evaluate detection guidelines and provide improvement suggestions for automating the detection.

\emph{Vulnerability databases.}
Common Vulnerabilities and Exposures (CVE) (launched in 1999) is the largest and most updated vulnerability database. 
Maintained by the MITRE corporation, it provides a publicly available list of most common security vulnerabilities with unique identifiers.
Common Weakness enumeration (CWE) is a community-developed list of common software security weaknesses. 
CWE aids developers and security practitioners since it serves as a common language for describing security weaknesses in architecture and implementation. 
It also provides different mitigation and prevention techniques that could be used to eliminate weaknesses. 

\emph{Architectural bad smells and anti-patterns.}
The literature on architectural bad smells \cite{taibi2018definition,garcia2009toward} and anti-patterns \cite{mo2019architecture,bouhours2009bad,nafees2017vulnerability} collides with our work for what concerns the ambition to find and remove architectural issues that negatively impact system life-cycle properties (extensibility, maintainability, testability, etc.).
In contrast, our work investigates security design flaws, that is architectural design decisions that negatively impact the system security.
Mo et al. \cite{mo2019architecture} have recently developed an automated detection of 6 architectural anti-patterns and study their impact on error and change-proneness of the related files.
The authors analyze 100 industrial software projects with respect to the project structural information and revision history.
They find that there only a few distinct types of anti-patterns that occur in all the projects, where \textit{Unstable Interface} and \textit{Crossing} were by far the largest culprits of error and change-proneness.
In contrast to our paper, the authors detect the anti-patterns based on existing implementation.
Our work is focused on detecting security design flaws at the level of architectural models.
Bouhours et al. \cite{bouhours2009bad} introduce a catalog of `spoiled patterns' and use it to automatically detect their manifestation in software architecture models.
They develop a plug-in for the Neptune environment (UML, XML) which supports the instantiation of spoiled patterns and suggest model transformations for re-factoring the design models.
Nafees et al. \cite{nafees2017vulnerability} propose a new template for detecting architectural anti-patterns and a catalog of 12 Vulnerability Anti-Patterns (VAP) entries.
The authors also provide some examples of the proposed VAP entries.
Yet, the catalog has not yet been evaluated empirically.
Taibi and Lenarduzzi \cite{taibi2018definition} have conducted interviews with 72 developers to collect a catalog of 11 microservice-specific bad smells.
Besides the 11 smells, the authors put forward the importance of carefully analyzing the connections between microservices, especially the connections leading to private data and shared libraries.
Similar to this work, the purpose of compiling such a catalog is to help practitioners in the detection of bad architectural decisions.
Garcia et al. \cite{garcia2009toward} describe 4 architectural bad smells identified through an in-depth analysis of two industrial systems.
The architectural smells are described in detail and are accompanied with UML component diagrams, and a discussion of their impact on quality.

\emph{Knowledge-based Architectural Threat Analysis.}
Architectural threat analysis consists of techniques and methods that are used for systematically analyzing the attacker's profile vis-a-vis assets of value to organizations.
Such techniques are often performed on models representing the software architecture of a system.
The purpose of analyzing security threats at this stage is to ultimately identify security holes and plan for necessary security solutions.
Therefore, we consider existing literature that makes use of knowledge base (threat catalogs, vulnerability data bases, etc.) \cite{abe2013modeling,almorsy2013automated,berger2016automatically,shostack2014threat,deng2011privacy} to perform such analysis as related work.
We refer the interested reader to a systematic literature review \cite{tuma2018threat} for a more detailed list of knowledge-based threat analysis techniques.


%% file: sec/threats.tex
\section{Threats to Validity}
\label{sec:threats}
With respect to \textit{internal validity threats}, we consider the threat of using graduate and doctoral students as participants. 
Using students during empirical studies has been criticized in the past, as their background knowledge is not that of industrial practitioners.
However, studies have shown~\cite{runeson2003using, salman2015students} that the differences between the performance of professionals and graduate students are often limited.
To counter this threat we have made sure that the selected participants had sufficient background education to complete the tasks.
We also consider the threat of an unrepresentative sample size (in total seven participants).
In addition, the suggestions for improvement have not been empirically evaluated.

With respect to \textit{external threats} to validity, we consider the threat to generalizability of results.
This study is conducted only on one experimental subject, thus the results can not be generalized to other domains.
To counter this threat, we plan to conduct more studies, including security experts, and different experimental subjects.

%% file: sec/conclusions.tex
\section{Conclusion}
\label{sec:conclusion}

This paper proposes a catalog of security design flaws accompanied by inspection guidelines for their detection.
To evaluate our approach, we present two empirical studies with master students and doctoral candidates.
We conduct two experiments investigating the performance of manually applying the inspection guidelines in the context of analyzing a home monitoring system.
The main contributions of this work are three-fold: (i) a catalog of security design flaws, (ii) an empirical evaluation of the inspection guidelines with master students, and (iii) a replicated evaluation with doctoral students.
We also account for the shortcomings of the inspection guidelines and make suggestions for their improvement with respect to the generalization of guidelines, catalog re-organization, and format of documentation.
Our results show a relatively high performance ($92.6$) and productivity ($11.5 \ TP/h$).
On the other hand, we found that about half of the security design flaws go unnoticed (average recall is $50.4\%$).
We introduce an additional measure to investigate which guidelines posed problems to our participants.
We identify three type of problems the participants encountered.
First, some guidelines were not general enough to be useful for detecting security design flaws.
Second, several closed questions used for detection were overlapping, potentially resulting in loss of time.
Lastly, participants have expressed a dislike for the format of documenting the flaws.
We suggest simple improvements of the guidelines to support future automation.